\documentclass{article}
\usepackage{frascatiphys}
\usepackage{graphicx}
\begin{document}
\title{\bf {HOLOGRAPHIC THEORY OF GRAVITY AND COSMOLOGY}
}
\author{Y. Jack Ng\\
{\em Institute of Field Physics, Department of Physics \& Astronomy,}\\
{\em University of North Carolina, Chapel Hill, NC 27599-3255, USA} 
}
\maketitle
\baselineskip=11.6pt
\begin{abstract}

According to the holographic principle, 
the maximum amount of information stored in a 
region of space scales as
the area of its two-dimensional surface, like a hologram.  We show 
that the holographic principle
can be understood heuristically as originated from quantum fluctuations 
of spacetime.  Applied to
cosmology, this consideration leads to a dynamical cosmological constant 
$\Lambda$
of the observed magnitude, in agreement with the result obtained 
for the present and recent cosmic eras, by using
unimodular gravity and causal-set theory.
By generalizing the concept
of entropic gravity, we find a critical acceleration 
parameter related to $\Lambda$ in galactic dynamics,
and we construct a phenomenological model of dark matter which we call 
``modified 
dark matter" (MDM).  We provide successful 
observational tests of MDM at both the galactic and cluster scales.
We also discuss the possibility that the quanta of both
dark energy and dark matter obey 
the quantum Boltzmann statistics or 
infinite statistics
as described by a curious average of the bosonic and 
fermionic algebras.

\end{abstract}
\baselineskip=14pt

\section{\bf {Introduction and Summary}}

In Vulcano 2004, in the talk "Space-time fluctuations," I discussed some 
aspects of "space-time foam" -- a foamy structure of spacetime arising from 
quantum fluctuations.\cite{ng04} To examine how large the fluctuations are, I 
considered a gedankan experiment in which a light signal is sent from a clock to 
a mirror (at a distance $l$ away) and back to the clock in a timing experiment 
to measure $l$.  From the jiggling of the clock's position alone, the Heisenberg 
uncertainty principle yields $\delta l^2 \stackrel{>}{\sim} \frac{\hbar l}{mc}$, 
where $m$ is the mass of the clock. On the other hand, the clock must be large 
enough not to collapse into a black hole; this requires $\delta l 
\stackrel{>}{\sim} \frac{Gm}{c^2}$.  We conclude that the fluctuation of a 
distance $l$ scales as $\delta l \stackrel{>}{\sim} l^{1/3} l_P^{2/3}$ (where 
$l_P = \sqrt{\hbar G/ c^3}$ is the Planck length).\cite{ng94} I also 
showed 
that this scaling of $\delta l$ is what the holographic principle 
\cite{holography} demands.

The present talk is a continuation of the talk I gave twelve years ago.  I will
start (in Section 2) by rederiving this scaling of $\delta l$ by another
method \cite{llo04}
which can be generalized to the case of an expanding universe for which
a dyamical cosmological constant is shown to emerge, 
\cite{ng08} a result that was earlier obtained \cite{PRL} 
by a consideration (in Section 3) of unimodular 
gravity \cite{Bij} and Sorkin's causal-set theory. 
Then I will discuss my recent work with Ho and Minic, 
and later work also with Edmonds, Farrah and Takeuchi. We found it 
natural (see Section 4) to generalize Verlinde's formulation 
\cite{verlinde} of entropic gravity/gravitational thermodynamics 
to de-Sitter space with a positive cosmological constant.  
The result was a dark matter model which we call modified dark 
matter (MDM).\cite{HMN}  Recently we have successfuly tested 
MDM (see Section 5) with 30 galactic rotation curves
and a sample of 93 galactic clusters.\cite{Edm1}

The take-home message from this talk is this:  
It is possible that the dark 
sector (dark energy and dark matter) has its origin in quantum gravity. 
And if the scenario to be sketched in Section
6 is correct, then we can expect some rather novel particle phenomenology, 
for the quanta of the dark sector obey not the familiar Bose or
Fermi statistics, but an exotic statistics that goes by the name
infinite statistics \cite{DHR} or quantum Boltzmann statistics. 
\cite{plb,HMN}  

I would like to take this opportunity to make a disclaimer: 
In a recent paper
``New Constraints on Quantum Gravity from X-ray and Gamma-Ray
Observations" by Perlman et al. (ApJ. 805, 10 (2015)), it was 
claimed that detections of quasars at
TeV energies with ground-based Cherenkov
telescopes seem to have ruled out the holographic spacetime foam model
(with $\delta l$ scaling as $l^{1/3} l_P^{2/3}$).  But now I 
(one of the authors) believe 
this conclusion is conceivably premature when proper averaging is carried out
(though presently there is no formalism yet for carrying out such averages.) 

\section{\bf {Spacetime Foam and the Cosmological Constant $\Lambda$}}

We can rederive the scaling of $\delta l$ by another argument. 
Let us consider mapping
out the geometry of spacetime for a spherical volume of radius $l$ over the
amount of time $2l/c$ it takes light to cross the volume.\cite{llo04} 
One way to 
do this is to fill the space with clocks, exchanging
signals with the other clocks and measuring the signals' times of arrival. 
The total number of operations, including the ticks of the clocks and
the measurements of signals, is bounded by the Margolus-Levitin theorem 
which stipulates that the rate of operations cannot exceed the amount 
of energy 
$E$ that is available for the operation divided by $\pi \hbar/2$.  
This theorem, combined with the
bound on the total mass of the clocks to prevent black hole formation, implies 
that the total number of operations that can occur in this spacetime volume is 
no bigger than
$2 (l/l_P)^2 / \pi$.
To maximize spatial resolution, each clock must tick
only once during the entire time period.  If we regard the operations
as partitioning the spacetime volume into ``cells", then 
there are $\stackrel{<}{\sim} l^2 / l_P^2$ cells, and,
on the average, each cell
occupies a spatial volume $\stackrel{>}{\sim} l^3 / (l^2 / l_P^2)
= l l_P^2$, yielding an average separation between neighboring
cells $ \stackrel{>}{\sim} l^{1/3} l_P^{2/3}$.
\cite{ng08}  This spatial separation can be interpreted as the average minimum 
uncertainty in the
measurement of a distance $l$, that is, $\delta l \stackrel{>}{\sim} l^{1/3}
l_P^{2/3}$.

It is straightforward to generalize \cite{ng08,plb} the above discussion for a 
static spacetime region with low spatial curvature to the
case of an expanding universe by the substitution 
of $l$ by $H^{-1}$ in the expressions for energy and entropy densities, 
where $H$ is the Hubble parameter. (Henceforth we 
adopt $c=1=\hbar$ for convenience unless stated otherwise for clarity.)  
Applied to cosmology, the above argument leads to the prediction
that (1) the cosmic energy density has the critical value $\rho \sim 
(H/l_P)^2$, and (2) the universe
of Hubble size $R_H$ contains $ I \sim (R_H/l_p)^2$ bits of 
information.  It follows that the average energy carried by 
each particle/bit is $\rho R_H^3/I \sim R_H^{-1}$.
Such long-wavelength constituents of dark energy give rise to
a more or less spatially uniform distribution of cosmic energy density and
act as a dynamical cosmological constant with the observed small but 
nonzero value $\Lambda \sim 3 H^2$.

\section{\bf Quantum (Generalized Unimodular) Gravity and (Dynamical) 
$\Lambda$}
 
The dynamical cosmological constant we have just obtained will be seen
to play an important role in our subsequent discussions.  So let us 
``rederive" it by using another method based on quantum gravity.
The idea makes use of the theory of
unimodular gravity\cite{Bij,PRL}, or better yet, the
generalized action given by 
$
S_{unimod} = - (16 \pi G)^{-1} \int [ \sqrt{g} (R + 2 \Lambda) - 2
\Lambda
\partial_\mu {\mathcal T}^\mu](d^3x)dt.
$
In this theory,
$\Lambda / G$ plays the role of
``momentum" conjugate to the ``coordinate" $\int d^3x {\mathcal T}_0$ which
can be identified
as the spacetime volume $V$.  Hence 
the fluctuations of $\Lambda /G$ and $V$ obey a 
quantum uncertainty principle,
$
\delta V \! \delta \Lambda/G \sim 1.
$

Next we borrow an argument due to Sorkin,
drawn from the causal-set theory, which
stipulates that continous geometries in classical gravity should be
replaced by "causal-sets", the discrete substratum of spacetime.
In the framework of the causal-set theory, the
fluctuation in the number of elements $N$ making up the set is of the
Poisson type, i.e., $\delta N \sim \sqrt{N}$.  For a causal set, the
spacetime volume $V$ becomes $l_P^4 N$.  It follows that
$
\delta V \sim l_P^4\delta N \sim l_P^4 \sqrt{N}
\sim l_P^2\sqrt{V} = G \sqrt{V},
$
and hence $\delta \Lambda \sim V^{-1/2}$.
By following an argument due to Baum and Hawking, 
we can show \cite{PRL} that, in the
framework of unimodular gravity, for the present and recent cosmic
eras (with matter fields being essentially in their ground states),
$\Lambda$ vanishes to the lowest order of
approximation and that its first order correction is positive.
Thus we conclude that $\Lambda$ is positive with a magnitude of
$V^{-1/2} \sim R_H^{-2}$, 
contributing a cosmic energy density $\rho$ given by:
$
\rho \! \sim \!  \frac{1}{l_P^2 R_H^2},
$
which is of the order of the critical density as observed!

\section{\bf {From $\Lambda$ to Modified Dark Matter (MDM)}}

The dynamical cosmological constant (originated from quantum fluctuations of 
spacetime) can now be
shown to give rise to a critical acceleration
parameter in galactic dynamics.  The argument \cite{HMN} is based on
a simple generalization of Verlinde's recent proposal of entropic gravity 
\cite{verlinde} for $\Lambda = 0$ to the case of de-Sitter space 
with positive $\Lambda$.  Let us first review Verlinde's derivation 
of Newton's second law $\vec{F} = m \vec{a}$.  
Consider a particle with mass $m$ approaching a holographic screen
at temperature $T$.  Using the first law of thermodynamics to introduce the 
concept of entropic force
$
F = T \frac{\Delta S}{\Delta x},
$
and invoking Bekenstein's original arguments 
concerning the entropy $S$ of black holes,
$\Delta S = 2\pi k_B \frac{mc}{\hbar} \Delta x$,
Verlinde gets $ F = 2\pi k_B \frac{mc}{\hbar} T$. With the aid of
the formula for the Unruh temperature, $k_B T = \frac{\hbar a}{2 \pi c}$,
associated with a uniformly accelerating (Rindler) observer, Verlinde
obtains $\vec{F} = m \vec{a}$.
Now in a de-Sitter space with positive cosmological 
constant $\Lambda$ for an accelerating universe like ours, 
the net Unruh-Hawking temperature, 
as measured by a non-inertial observer with 
acceleration $a$ relative to an inertial observer, is 
$\tilde{T} = \frac{\hbar \tilde{a}}{2\pi k_B c}$ 
with $\tilde{a} \equiv \sqrt{a^2+a_0^2} - a_0$, \cite{deser}
where $a_0 \equiv \sqrt{\Lambda / 3}$.  Hence the
entropic force (in de-Sitter space) is given by the replacement of $T$ and 
$a$ by  $\tilde{T}$ and $\tilde{a}$ respectively, leading to
$F =  m [\sqrt{a^2+a_0^2}-a_0]$.
For $ a \gg a_0$, we have $F/m \approx a$ which gives $a = a_N \equiv 
GM/r^2$, the Newtonian acceleration.  But for 
$a \ll a_0$, $F \approx m \frac{a^2}{2\,a_0} = m v^2/r$ for 
circular motions, so
the observed flat galactic rotation curves ($v$ being independent of $r$) now 
require
$ a \approx \left(2 a_N \,a_0^3/\pi \, \right)^{\frac14}$.
But that means $F \approx m \sqrt{a_N a_c}\,$,
the modified Newtonian dynamics (MoND)
scaling \cite{mond}, proposed by Milgrom. 
Thus, we have recovered MoND with the correct
magnitude for the critical galactic acceleration parameter 
$a_c = a_0/ (2 \pi) \approx c H / (2 \pi) \sim 10^{-8} cm/s^2$. 
As a bonus, we have also recovered the observed 
Tully-Fisher relation ($v^4 \propto M$).

Next we \cite{HMN} can follow the 
second half of Verlinde's argument \cite{verlinde} to 
generalize Newton's law of gravity $a= G M /r^2$.  
The end result is given by
$\tilde{a}  = G\,\tilde{M} /r^2$,
where $\tilde{M} = M + M_d$ 
represents the \emph{total} mass enclosed within 
the volume $V = 4 \pi r^3 / 3$, 
with $M_d$ being some unknown mass, i.e., dark
matter.  For $a \gg a_0$, consistency with the Newtonian force law $a \approx 
a_N$ implies $M_d \approx 0$.  But
for $a \ll a_0$, consistency with the condition 
$a \approx \left(  2 a_N \,a_0^3 / \pi \right)^{\frac14}$ 
requires $M_d \approx \frac{1}{\pi} \left(\,\frac{a_0}{a}\,\right)^2\, M
\sim (\sqrt{\Lambda}/G)^{1/2}M^{1/2}r$. 
(Note that $M_d$ depends on $\Lambda$ and $M$.) 
Actually we can interpret the MoNDian force law as 
a manifestation of dark matter. That explains why initially we called
our model ``MoNDian dark matter" \cite{HMN}, which, to some people, 
sounds like an oxymoron; so
now we call it ``modified dark matter." \cite{Edm1}

\section{\bf Observational Tests of MDM}

In order to test MDM with galactic rotation curves, we fit computed rotation 
curves to a selected sample of Ursa Major galaxies given in \cite{Sanders98}, 
using the mass-to-light 
ratio $M/L$ as our {\it only} fitting parameter.
For the CDM fits, we use 
the Navarro, Frenk \& White density profile, employing {\it 
three} free parameters (one of which is
the mass-to-light ratio.)  We find that  
both models fit the data well (and more or less equally well)!
But while the MDM fits use only 1 free parameter,
for the CDM fits one needs 3 free parameters.  Thus the MDM model is a 
more economical model than CDM in fitting data at the galactic scale.
As for dark matter density, the profiles 
predicted by MDM and CDM agree well in the asymptotic (large $R$) regime. 
See Ref. \cite{Edm1} for details.

To test MDM with astronomical observations at a larger scale, we \cite{Edm1}
compare dynamical and observed masses in a large sample of galactic clusters
studied by Sanders \cite{Sanders1999} using the compilation
by White, Jones, and Forman. 
Sanders \cite{Sanders1999} studied the virial discrepancy (i.e.,
the discrepancy between the observed mass and the dynamical mass) in the 
contexts of Newtonian dynamics and MoND.
He found the well-known discrepancy
between the Newtonian dynamical mass ($M_{\textrm{N} }$) and the observed mass
($M_{\textrm{obs} }$):
$
\left \langle \, \frac{M_\textrm{N}}{ M_{\textrm{obs} }} \,\right \rangle
\approx 4.4\,.
$
And for the sample clusters, he found $\langle M_{\textrm{MoND}} /
M_{\textrm{obs}} \rangle \approx 2.1.$

We \cite{Edm1} have adapted  Sanders'
approach to the case of MDM.
Noting that the argument used in Section 4 does allow
$M_d$ to include a term of the form $\xi \left(\,\frac{a_0}{a}\,\right)\, M$ 
with an undetermined universal parameter $\xi$, we (in some unpublished 
work) have decided to use a more general profile of the form
$
M_d = \left[\,\xi\,\left(\,\frac{a_0}{a}\,\right)+\frac{1}{\pi}\,
\left(\,\frac{a_0}{a}\,\right)^2\,\right] \, M\,.
$
For $\xi \approx 0.5$, we get
$
\left \langle \, \frac{M_\textrm{MDM}}{ M_{\textrm{obs} }} \,\right \rangle
\approx 1.0\,.$
(As an aside, we have refit the galaxy rotation curves using $\xi = 0.5$ and 
have found equally good fits.)
Thus the virial discrepancy is eliminated in the context of MDM! 
At the cluster scale, MDM is superior to MoND.

\section{\bf {The Dark Sector and Infinite Statistics}}

What is the essential difference between ordinary matter and
dark energy from our perspective?
To find that out, let us recall our discussion in Section 2, and liken
the quanta of dark energy to
a perfect gas of $N$ particles obeying Boltzmann statistics
at temperature $T$ in a volume $V$.  For the
problem at hand, as the lowest-order approximation, we can neglect the
contributions from matter and radiation to the cosmic 
energy density for the recent and present eras.  
Thus let us take $V \sim R_H^3$, $T \sim R_H^{-1}$, and $N \sim
(R_H/ l_P)^2$. A standard calculation (for the relativistic case) yields the
partition function $Z_N = (N!)^{-1} (V / \lambda^3)^N$, where
$\lambda = (\pi)^{2/3} /T$,
and we get, for the entropy of the system,
$
S = - ( \partial (-T ln Z_N) / \partial T)_{V,N} 
= N [ln (V / N \lambda^3) + 5/2].
$

The important point to note is that, since $V \sim \lambda^3$, the entropy
$S$ becomes nonsensically negative unless $ N \sim
1$ which is equally nonsensical because $N$ should not be too different from
$(R_H/l_P)^2 \gg 1$.
But the solution \cite{plb} 
is obvious: the $N$ inside the log of $S$ somehow
must be absent.  
That is the case if 
the Gibbs $1/N!$ factor
is absent from the partition function $Z_N$,
implying that the ``particles" are
distinguishable and nonidentical!

Now the only known consistent statistics in greater than two space
dimensions
without the Gibbs factor 
is infinite statistics (sometimes called
``quantum Boltzmann statistics") \cite{DHR}.  Thus 
the ``particles" constituting dark energy obey infinite statistics,
instead of the familiar Fermi or Bose statistics. \cite{plb}

To show that the quanta of MDM also obey
this exotic statistics, we \cite{HMN} first reformulate MoND via an effective 
gravitational dielectric medium, motivated by the analogy \cite{dielectric} 
between Coulomb's law in a dielectric medium and Milgrom's law for MoND.
Ho, Minic and I then find that MoNDian force law is recovered if the 
quanta
of MDM obey infinite statistics.

What is infinite statistics?  Succinctly, a Fock realization of infinite
statistics is provided by the commutation relations of
the oscillators:
$a_k a^{\dagger}_l = \delta_{kl}$. 
Curiously a theory of particles
obeying infinite statistics cannot be local \cite{DHR}.  
But the TCP theorem and cluster
decomposition have been shown to hold despite the lack of locality
\cite{DHR}.  Actually this lack of locality 
is not unexpected.  After all, non-locality is also present in 
holographic theories, and  
the holographic principle is 
an important ingredient in the formulation of quantum gravity.
Infinite statistics and quantum gravity appear to fit together
nicely, and non-locality seems to be a common feature of both of them.
\cite{plb}  Perhaps it is the extended nature of the dark quanta that
connects them to such global aspects of space-time as the Hubble 
parameter and the cosmological constant.\\

\noindent
{\bf Acknowledgment}

This talk is partly based on work done in collaboration with  
H. van Dam, J.J. van der Bij,
S. Lloyd, M. Arzano, T. Kephart, C. M. Ho, D. Minic, D. Edmonds,
D. Farrah, and T. Takeuchi.  I thank them all.
The work reported here
was supported in part by the US Department of Energy, the Bahnson
Fund and the Kenan Professorship Research Fund of UNC-CH.
\end{document}